\documentclass[prb,letterpaper,aps,floatfix,twocolumn]{revtex4-1}
\usepackage{graphicx}
\usepackage{amsmath}
\usepackage{slashed}
\usepackage{extarrows}
\newcommand{\beq}{\begin{equation}}
\newcommand{\eeq}{\end{equation}}
\newcommand{\bk}{{{\bf{k}}}}

\newcommand{\br}{{{\bf{r}}}}

\newcommand{\bq}{{\bf{q}}}

\newcommand{\bp}{{\bf{p}}}

\newcommand{\beqa}{\begin{eqnarray}}
\newcommand{\eeqa}{\end{eqnarray}}

\begin{document}
\title{Impurity scattering and Friedel oscillations in mono-layer black phosphorus}
\author{Y. L. Zou$^1$, J. T. Song$^2$, C. X. Bai$^1$}
\author {K. Chang$^1$}
\email{kchang@semi.ac.cn}
\affiliation{$^1$SKLSM, Institue of Semiconductors, Chinese Academy of Sciences, P. O. Box 912, Beijing 100083, China\\
$^2$Department of Physics and Hebei Advanced Thin Film Laboratory, Hebei Normal University, Hebei 050024, China}

\date{\today}
\begin{abstract}
We study the effect of impurity scattering effect in black phosphorurene (BP) in this work. For single impurity, we calculate impurity induced local density of states (LDOS) in momentum space numerically based on tight-binding Hamiltonian. In real space, we calculate LDOS and Friedel oscillation analytically. LDOS shows strong anisotropy in BP. Many impurities in BP are investigated using $T$-matrix approximation when the density is low. Midgap states appear in band gap with peaks in DOS. The peaks of midgap states are dependent on impurity potential. For finite positive potential, the impurity tends to bind negative charge carriers and vise versa. The infinite impurity potential problem is related to chiral symmetry in BP.
\end{abstract}
\maketitle
\section{Introduction}
\label{sec:1}

Graphene has remarkable electronic, optical and mechanical properties and shows promising applications in electronic devices.~\cite{Neto09} Since the successful procduction of grpahene in 2004,~\cite{Novoselov04} there have been many studies on graphene both experimentally and theoretically.~\cite{Neto09} However, due to its gapless spectrum, graphene is not a good candidate for on-off devices. It is desirable to find 2-dimensional materials with a tunable gap which can be utilized as on-off devices. A promising candidate reported in recent years is mono-layer BP which has a direct band gap.~\cite{Zhang14}

 Mono-layer BP has been fabricated in lab using exfoliation method.~\cite{Zhang14} Since then, more and more works has focused on single layer and few layer BP.~\cite{Neto2014,Ezawa2014,Guinea2014,Katsnelson2015,Chang2015b,Low2015,Chang2015} Bulk BP has a band gap of $0.31-0.35$ eV, and single layer BP has a band gap of about $1.5$ eV.\cite{Zhang14} Mono-layer BP has large mobility as well as a band gap which makes it a promising material in electronic devices. It has different masses along armchair and zigzag direction respectively. This anisotropy has great impact on its transport and optical properties. For example, Tony Low et al~\cite{Guinea2014} studied the plasmons in BP and found that the plasmon excitations have different dispersion along different directions.

 Impurity is introduced in processes of fabricating materials. So it is important to understand impurity effects in mono-layer BP because impurity can affect properties of devices remarkably.
  Impurities could be introduced in many ways. Substrates provide a source of impurities and the adsorbed atoms on BP or missing atoms in BP can induce vacancies and so on.~\cite{Sarma2011} Impurities provide scattering centers to carriers and are the main contribution to lifetime of carriers in the low-temperature limit. We mainly focus on short-ranged impurities in this paper.~\cite{Sarma2011} To understand impurities effects in BP will provide useful information about this new fabricated material.

To understand how impurity affects carriers in BP, it is necessary to study the density of states (DOS) of carriers.
The electronic DOS has many characteristics showing presence of impurities. There are many physical properties closely related to DOS of carriers, such as electric conductivity, optical conductivity, scanning-tunneling microscope (STM) images. The DOS of carriers is also related to the polarization of the system which modifies the electronic screening.

For a single impurity, the Fourier transform scanning tunneling spectroscopy (FT-STS) which is Fourier transformation of local density of states (FT-LDOS) can reveal many informations about carriers~\cite{Lee2003,Bena2005,Bena2008,MacDonald2008,Hu2009}. The FT-STS are interference patterns which originate from interference of incoming waves and outgoing waves scattered by the impurity. The scattering occurs on contours of constant energy and momentum is transferred from the impurity to carriers, so the amplitudes in FT-STS can reveal what kind of scattering can happen and also the properties of the impurity. For example, the FT-STS in monolayer graphene and bilayer graphene show different signatures which can be used to distinguish monolayer and bilayer graphene.~\cite{Bena2008} For surface states of 3-dimensional topological insulator, the backscattering is forbidden. So in experiment FT-STS amplitude is weak for the backscattering which makes it a strong evidence for topological insulator.~\cite{Hu2009} Therefore, FT-STS is a useful method in experiment to investigate material properties.

In real space, the disturbance of the impurity relocates electrons, and the amplitude of LDOS oscillates and decays away from the position of the impurity. So the electron density at position $\br$ which is obtained by summing LDOS at $\br$ up to Fermi energy will oscillate and decay in real space. It is called Friedel oscillation (FO). In graphene, FO has been studied in a series of papers.~\cite{Cheianov2006,Virosztek2010} FO decays as $r^{-3}$ in graphene, the oscillation wave vector is $2k_F$ with $k_F$ the Fermi surface wave vector.

For a finite but small density of impurities, peaks can show up in DOS. The states associated with these peaks are called midgap states because they appear in the band gap.~\cite{Katsnelson2015} In graphene, the midgap states appear at Dirac point where the DOS is zero.~\cite{Pereira2008} These midgap states can be viewed as bound states attracted by impurity potential.       

In this work, we first study FT-STS in BP which is caused by a single impurity, the numerical results are based on four band tight-binding Hamiltonian in BP using $T$-matrix.~\cite{Wehling2009,Wehling2007,Balatsky2010,Balatsky2006,Sarma2008,Peres2006,Pereira2008} This problem has been investigated in graphene and in topological insulator,~\cite{Hu2009,Bena2005} but the strong anisotropy in BP has strong influence on FT-STS and it is useful to study these anisotropy related effects in BP. We also investigate FO in real space based on two-band model of BP.~\cite{Ezawa2014} For the case of many impurities, we focus on short-ranged and low density impurities. We present calculations of DOS in BP using both $T$-matrix approximation (full Born approximation) and self-consistent $T$-matrix approximation (full self-consistent Born approximation).~\cite{Peres2006,Sarma2008} $T$-matrix is exact for single impurity and is a good approximation for many impurities as long as the density is low and impurity is short ranged.~\cite{Peres2006}

The rest of the paper is organized as follows: In sec. \ref{sec:tmatrix} we introduce the T-matrix method and also the self-consistent $T$-matrix method. In sec. \ref{sec:FT-STS} we calculate FT-STS in BP induced by a single impurity, sec. \ref{sec:FO} gives the Friedel oscillation in BP. Sec. \ref{sec:dos} deals with the low density of impurities using both $T$-matrix and self-consistent $T$-matrix method. Finally, we conclude in sec. \ref{sec:discuss}.


\section{Hamiltonian of BP and $T$-matrix}
\label{sec:tmatrix}

\begin{figure}
  \includegraphics[width=11cm]{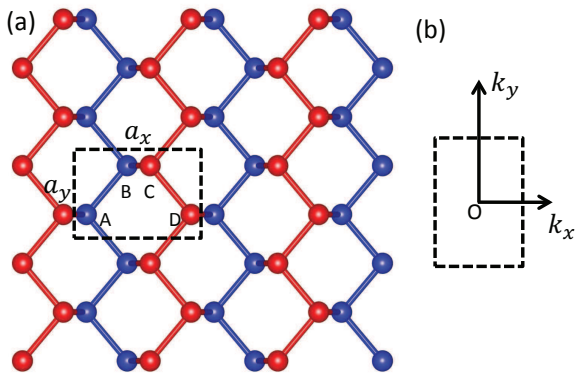}\\
  \caption{Lattice structure of monolayer black phosphorene. (a) Red (blue) atoms represent the upper (lower) layer, there are four atoms in a unit cell which are labeled A, B, C and D. The unit cell size along $x$ direction is $a_x=4.43$ \AA, $y$ direction is $a_y=3.27$ \AA. (b) The first Brillouin zone (BZ) of BP.}\label{fig:1}
\end{figure}

We start with our calculation using the tight-binding model of monolayer BP. BP has the same structure as graphene but with the atoms puckered. There are four phosphorus atoms in the unit cell of BP shown in figure \ref{fig:1}, the four-band Hamiltonian is
\beqa
\label{eq:1}
    H(\bk)=\left(
        \begin{array}{cccc}
          0 & f_{AB} & f_{AC} & f_{AD}\\
          f_{AB}^* & 0 & f_{BC} & f_{BD}\\
          f_{AC}^* & f_{BC}^* & 0 & f_{CD}\\
          f_{AD}^* & f_{BD}^* & f_{CD}^* & 0 \\
        \end{array}
      \right)
\eeqa

where
\begin{eqnarray}\label{eq:2}
f_{AB}&=&t_1(1+e^{-ik_ya_y})+t_3(e^{-ik_xa_x}+e^{-ik_xa_x-ik_ya_y}),  \nonumber \\
f_{AC}&=&t_4(1+e^{-ik_ya_y}+e^{-ik_xa_x}+e^{-ik_xa_x-ik_ya_y}),\nonumber\\
f_{AD}&=&t_2e^{-ik_xa_x}+t_5, \nonumber \\
f_{BC}&=&t_2+t_5e^{-ik_xa_x},\nonumber\\
f_{BD}&=&t_4(1+e^{ik_ya_y}+e^{-ik_xa_x}+e^{-ik_xa_x+ik_ya_y}), \nonumber \\
f_{CD}&=&t_1(1+e^{ik_ya_y})+t_3(e^{-ik_xa_x}+e^{-ik_xa_x+ik_ya_y})
\end{eqnarray}
with $a_x$, $a_y$ given in figure \ref{fig:1}, A, B, C and D denote the four atoms in unit cell. The tight-binding parameters read as $t_1=-1.220$ eV, $t_2=3.665$ eV, $t_3=-0.205$ eV, $t_4=-0.105$ eV, $t_5=-0.055$ eV.~\cite{Katsnelson2015b}

In this work we assume the impurity is short ranged and can be modeled as delta potential. In the vicinity of Dirac point in graphene, long-ranged Coulomb impurity is the main source of scattering,~\cite{Sarma2011} while in BP Coulomb impurity can be treated as short-ranged because of screening. So we mainly focus on short-ranged impurity. The impurity potential written in $\bk$ space is assumed to be
\begin{equation}\label{eq:4}
    U=u\left(
         \begin{array}{cccc}
           1 & 0 & 0 & 0 \\
           0 & 0 & 0 & 0 \\
           0 & 0 & 0 & 0 \\
           0 & 0 & 0 & 0 \\
         \end{array}
       \right),
\end{equation}
where $u$ is constant in $\bk$ space. The only nonzero element of $U$ at first row and first column means the impurity is near or resides on the single A atom.
We take $u\rightarrow\infty$ for vacancies. In the single impurity problem, we fix the position of the impurity in the first unit cell and take the impurity site as origin in real space. In the case of low density concentration of impurities, we fix the number of impurities and their positions are randomly distributed in the system. The final results take the average of all possible configurations of impurities. The Born approximation is often introduced in the calculation if $u$ is small and the impurity is short-ranged, but when bound state is formed near the impurity, the Born approximation is not justified. In stead, $T$-matrix should be introduced in calculation when bound states are formed.~\cite{Balatsky2006} The $T$-matrix is exact for single impurity for any value of $u$. For low density concentration of impurities or vacancies, the $T$-matrix approximation is also a good approximation.~\cite{Sarma2008} We also calculate the DOS using self-consistent $T$-matrix approximation and find that they do not make big differences.

For a single impurity, the LDOS is obtained by taking the imaginary part of full Green function. The full Green function is carried out using $T$-matrix and Matsubara frequency~\cite{Sarma2008}
\begin{equation}
\label{eq:5}
\begin{aligned}
    G(\bk_1,\bk_2,i\omega_n)=&G^0(\bk_1-\bk_2,i\omega_n)\\
    &+\sum_{\bk^\prime}G^0(\bk_1,i\omega_n)T_{\textrm{imp}}(\bk_1,\bk_2,i\omega_n)G^0(\bk_2,i\omega_n)
\end{aligned}
\end{equation}
written in momentum space,
where the $T_{\textrm{imp}}$ matrix satisfies the self-consistent equation
\beqa\label{eq:4}
    T_{\textrm{imp}}(\bk_1,\bk_2,i\omega_n)&=&U(\bk_1,\bk_2)\nonumber\\
    &+&U(\bk_1,\bk^\prime)G^0(\bk^\prime,i\omega)T_{\textrm{imp}}(\bk^\prime,\bk_2,i\omega_n),\nonumber\\
\eeqa
and the zeroth Green function in momentum space is
\begin{equation}\label{eq:6}
    G^0(\bk,i\omega_n)=[i\omega_n-H(\mathbf{k})]^{-1}\overset{i\omega_n\rightarrow \omega+i\eta}=[\omega+i\eta-H(\mathbf{k})]^{-1}
\end{equation}
where $\eta$ is set to $0.01$ eV in our numerical calculation.

Since BP has four atoms in a unit cell, the above equations are $4\times4$ matrix equation.
So the $T$ matrix can be obtained as
\begin{equation}\label{eq:7}
    T_{\textrm{imp}}(i\omega_n)=\frac{V}{I-V/N\sum_{\bk\in \textrm{BZ}}G^0(\bk,i\omega_n)},
\end{equation}
where the summation is over the first BZ, $I$ is $4\times4$ identity matrix and $N$ is the number of unit cells of BP. In the unitary limit ($u\rightarrow\infty$) the $T$ matrix reduces to
\begin{equation}\label{eq:8}
   T_{\textrm{imp}}(i\omega)=-\left[\overline{G}^0_{AA}(i\omega)\right]^{-1}\left(
                                                                          \begin{array}{cccc}
                                                                            1 & 0 & 0 & 0 \\
                                                                            0 & 0 & 0 & 0 \\
                                                                            0 & 0 & 0 & 0 \\
                                                                            0 & 0 & 0 & 0 \\
                                                                          \end{array}
                                                                        \right),
\end{equation}
where
\begin{equation}\label{eq:9}
    \overline{G}^0_{AA}(i\omega_n)=\frac{1}{N}\sum_{\mathrm{\mathbf{k}}\in \mathrm{BZ}}G^0_{AA}(\bk,i\omega_n).
\end{equation}

The LDOS induced by a single impurity located at the origin is
\begin{equation}\label{eq:10}
    \delta\rho(\bq,i\omega_n)=-\frac{1}{N\pi}\sum_{\bk\in \textrm{BZ}}\left[\delta G(\bk,\bk+\bq)-\delta G^*(\bk,\bk+\bq)\right],
\end{equation}
where
\begin{equation}\label{eq:11}
    \delta G(\bk,\bk+\bq)= G(\bk,\bk+\bq)-G^0(\bk,\bk+\bq).
\end{equation}

For many impurities, we take the small density into account, i.e. small $n_i=N_i/N$, where $N_i$ is the number of impurities. We average over all possible positions of impurities and approximate the Green function of carriers to first order in $n_i$ using $T$-matrix approximation. The Green function is
\begin{equation}\label{eq:12}
    G(\mathrm{\mathbf{k}},i\omega_n)=G^0(\bk,i\omega_n)+G^0(\mathrm{\mathbf{k}},i\omega_n)T(i\omega_n)G(\bk,i\omega_n),
\end{equation}
which can be solved as
\begin{equation}\label{eq:13}
    G(\bk,i\omega_n)=[(G^0(\bk,i\omega_n))^{-1}-T(i\omega_n)]^{-1}.
\end{equation}

In the above equations, $T$-matrix is $T(i\omega_n)=Un_i[1-U\overline{G}^0(i\omega_n)]^{-1}$. If we use the self-consistent $T$-matrix approach, $T$-matrix becomes $T(i\omega_n)=Un_i[1-U\overline{G}(i\omega_n)]^{-1}$, where $\overline{G}(i\omega_n)=1/N\sum_\bk G(\bk,i\omega_n)$. The difference between $T$-matrix approximation and self-consistent $T$-matrix approximation is that $T$-matrix uses the zeroth-order Green function while the latter uses the full Green function in calculating $T(i\omega_n)$.

\section{FT-STS for a single impurity}
\label{sec:FT-STS}

\begin{figure}
  \includegraphics[width=9cm]{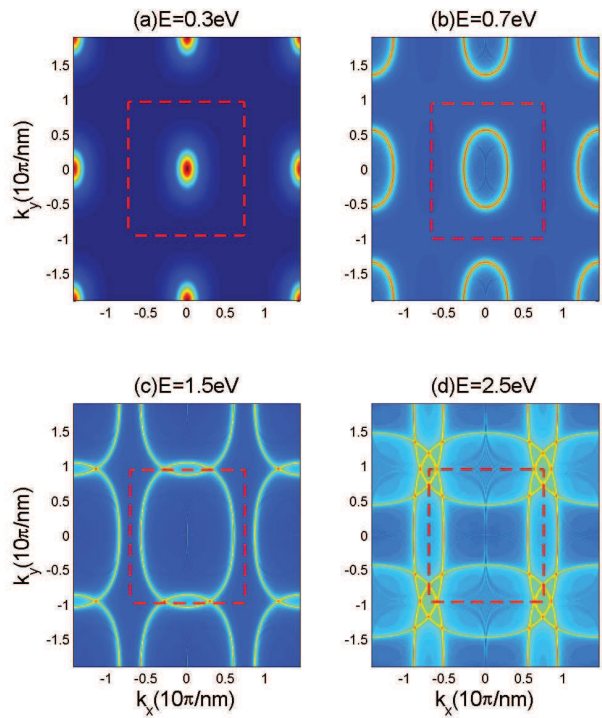}\\
  \caption{FT-LDOS in BP with (a) $E=0.3$ eV, (b) $E=0.7$ eV, (c) $E=1.5$ eV and (d) $E=2.5$ eV. The red dashed rectangle denotes first BZ. In the calculation of FT-LDOS, we set $u=2$ eV. The largest part in LDOS is elliptic contours.}\label{fig:ft-sts}
\end{figure}

  It is necessary at beginning to clarify some issues relating to experiments and conditions under which our approach is appropriate. FT-STS is related to Fourier transformed LDOS $\rho(\br_i,i\omega_n)$ with $\br_i$ the position of unit cell. The experimental situation is complicated. If the tip of STM has high resolution, then the LDOS related to the $4$ atoms can be detected respectively. If the resolution is low, the LDOS detected may be $\sum_{X=A,B,C,D}\rho_{XX}(\br_i)$. In our work we calculate $\rho_{AA}(\br_i)$ for simplicity. In the zero temperature limit, it is assumed that all other interactions, inelastic scattering can be incorporated into the broadening for simplicity. The reason is that in the low temperature limit, inelastic scattering is suppressed, the electron-electron interaction is screened and Landau's Fermi liquids picture is valid.

The FT-LDOS is equation \eqref{eq:10} which is a convolution integral essentially. The steps in numerical calculation based on tight-binding model are as follows: The zeroth order Green function in $k$-space at right hand of equation \eqref{eq:5} is Fourier transformed to real space, and we obtain $G^0(E+i\eta,\br_i)$ and $G^0(E+i\eta,-\br_i)$ respectively, then Fourier transform their products back to $\bk$-space including appropriate coefficients.  \\



 Before the presentation of numerical results, we discuss how the scattering process influence the interference pattern in FT-STS and do some analytical calculations.  First, the energy is conserved because the impurity is time-independent, namely, scattering between states occurs on the same energy contour. Second, from equation \eqref{eq:5}, it is easily obtained that the dominant contribution to FT-STS is where energy $\omega$ is close to the two poles of the two Green function simultaneously. The associated quantity is also called joint DOS.~\cite{Lee2003} Following T. Pereg-Barnea et al,~\cite{MacDonald2008} we use the $\bk\cdot\bp$ Hamiltonian and consider $E$ lying in conduction band taking into interband coupling as renormalzation of conduction effective masses, the Green function to first order in $u$ has the form


\beq
\label{eq:14}
    \delta G(\bq,i\omega)=\frac{1}{N}\sum_{\bk}[G^0_{AA}(i\omega,\bk)uG^0_{AA}(i\omega,\bk+\bq)],
\eeq
transforming sum into integral and using Feynman's parameter, changing variable $k_x\rightarrow\sqrt{\eta_c^\prime}k_x$, $k_y\rightarrow\sqrt{\nu_c}k_y$ and defining $p_x=\sqrt{\eta_c^\prime}q_x$, $p_y=\sqrt{\nu_c}q_y$, it becomes

\beqa
\label{eq:15}
\delta G(\bq,i\omega)
&=&\frac{A_c}{(2\pi)^2}\frac{u}{\sqrt{\eta_c^\prime\nu_c}}\int d^2\bk \frac{1}{i\omega- \bk^2}\frac{1}{i\omega-(\bk+\bp)^2}\nonumber\\
&=&\frac{A_c}{4\pi^2}\frac{u}{\sqrt{\eta_c\nu_c^\prime}}\int_0^1 dx\int d^2\bk \frac{1}{[i\omega-k^2-x(1-x)p^2]^2}\nonumber\\
&=&-\frac{1}{\pi}\frac{u A_c}{\sqrt{\eta_c^\prime\nu_c}p^2}\frac{1}{\sqrt{4\omega/p^2-1}}\arctan{\frac{1}{\sqrt{4\omega/p^2-1}}},\nonumber\\
\eeqa
where $A_c=a_xa_y$ is the area of a unit cell. So the FT-LDOS is
 \beq
 \label{eq:16}
 \delta\rho(\bq,E)=\frac{1}{\pi^2}\frac{u A_c}{\sqrt{\eta_c^\prime\nu_c}p^2}\rm{Im}\mathcal{F}(\frac{4E}{p^2})
 \eeq
 with
 \beq
 \mathcal{F}(z)=\frac{1}{\sqrt{z-1}}\arctan{\frac{1}{\sqrt{z-1}}}.
 \eeq
 The parameter $\eta_c^\prime$, $\nu_c$ relating to effective masses are given in the next section.


According to equation \eqref{eq:16}, the FT-STS is zero outside the contour $p^2=4E$, i.e. $\eta_c q_x^2+\nu_c q^2_y=4E$. This contour becomes a branch cut for $\mathcal{F}(\frac{4E}{p^2})$. In numerical calculation, the broadening of levels cause FT-LDOS in regions outside this contour nonzero. It is clear that the largest part lies on this contour.

The numerical result is shown in figure \ref{fig:ft-sts} with (a) $E=0.3$ eV, (b) $E=0.7$ eV, (c) $E=1.5$ eV and (d) $E=2.5$ eV. The scattering occurs on the contours of constant energy and the results show that large amplitude of FT-LDOS comes from scattering exchanging momentum $\bq=2\bk_E$ with $\bk_E$ corresponding to wave vector on energy contour of $E$. There is only intraband scattering in BP while in graphene scattering can happen between two nonequivalent $K$ points. However, BP shows strong anisotropy in FT-STS.
 The prominent feature is that the constructive interference occurs on elliptic contours which is consistent with equation \eqref{eq:16}. The resulting FT-LDOS in BP is different from graphene. In graphene, FT-LDOS has a circle around $\Gamma$ and six trigonal contour around six vertex point of BZ,~\cite{Bena2005} while in BP, the contour is around $\Gamma$ and its shape is elliptic due to different masses in two directions. This largest part of FT-LDOS in BP is backscattering where momentum transfer is two times Fermi wave vector, i.e. the scattering just reversed the wave vector.

 Note there are also small interference patterns inside the elliptic contours, these small parts of FT-LDOS is due to terms proportional to the product $\rm{Re}{\frac{1}{E+i\eta-\varepsilon(\bk)}}\rm{Im}{\frac{1}{E+i\eta-\varepsilon(\bk+\bq)}}$ with $\bk$ fixed around a point on energy contour and $\bq$ variable. This small contribution also exists in FT-LDOS in graphene, but the isotropy in energy dispersion makes this part is the same in all directions and it will not emerge. In BP, this small part has different values in different directions, and the large part emerges.
 The incoming waves interference with the outgoing waves after scattering which is called FO in real space. We will make detailed studies on it in next section. The strong anisotropy in FT-STS of BP makes the pattern more easily identified in STM experiments.

\section{$\bk\cdot\bp$ Hamiltonian and FO in BP}
\label{sec:FO}

 To investigate FO in BP, we use $2\times2$ $\mathbf{k}\cdot\mathbf{p}$ Hamiltonian of BP. The Hamiltonian can be described as~\cite{Neto2014,Ezawa2014}
\begin{equation}\label{eq:4.1}
    H_{kp}=\left(
        \begin{array}{cc}
          E_c+\eta_ck_x^2+\nu_ck_y^2 & \gamma k_x \\
          \gamma k_x & E_v-\eta_vk_x^2-\nu_vk_y^2 \\
        \end{array}
      \right),
\end{equation}
where subscript $c(v)$ label the conduction (valence) band,  $\eta_{c,v}$ and $\nu_{c,v}$ are related to the effective masses by $\eta_{(c,v)}=\hbar^2/2m_{(c,v)x}$, $\nu_{(c,v)}=\hbar^2/2m_{(c,v)y}$, the mass parameter are $m_{cx}=0.151m_e$, $m_{cy}=1.062m_e$, $m_{vx}=0.122m_e$, $m_{vy}=0.708m_e$ with $m_e$ the free electron mass. $E_c=0.34$ eV ($E_v=-1.018$ eV) is the conduction (valence) band edge, $\gamma=-5.231$ $\textrm{eV}\cdot\textrm{\AA}$ is interband coupling coefficient. This Hamiltonian can be obtained from tight-binding Hamiltonian by low-energy expansion to second order of $\bk$ near $\Gamma$ point based on $D_{2h}$ symmetry of BP.~\cite{Ezawa2014} The $\bk\cdot\bp$ Hamiltonian has strong anisotropic masses along the $x$ and $y$ directions and we will use it to calculate FO in BP.

%
%

We briefly describe how to obtained equation \eqref{eq:4.1} from tight-binding Hamiltonian \eqref{eq:1} here. The details can be found in references~\onlinecite{Katsnelson2015b,Ezawa2014}. The eigenvector of tight-binding Hamiltonian \eqref{eq:1} is given by $[\phi_A\phi_B\phi_C\phi_D]$. Using unitary transformation
\beq
\label{eq:4.2}
\frac{1}{\sqrt{2}}\left(
  \begin{array}{cccc}
    1 & 0 & 1 & 0 \\
    0 & 1 & 0 & 1 \\
    1 & 0 & 1 & 0 \\
    0 & 1 & 0 & 1 \\
  \end{array}
\right),
\eeq
the tight-binding Hamiltonian can be reduced to a block diagonal Hamiltonian with each block a $2\times2$ Hamiltonian of which the eigenvector is
\beq
\label{eq:4.3}
\Psi=
\frac{1}{\sqrt{2}}
\left(
  \begin{array}{c}
    \phi_A+\phi_C \\
    \phi_B+\phi_D \\
  \end{array}
\right).
\eeq
So BP can be described by a $2\times2$ Hamiltonian. After a rotation of the Pauli matrices $\tau_x\mapsto\tau_z$ followed by $\tau_y\mapsto\tau_x$,~\cite{Ezawa2014} this $2\times2$ Hamiltonian is transformed into equation \eqref{eq:4.1}. Under this rotation, the impurity potential U is transformed into
\beq
\label{eq:19}
U=u/\sqrt{2}\left(
       \begin{array}{cc}
         1 & -i \\
         i & 1 \\
       \end{array}
     \right).
\eeq

In this sec., we use Born approximation which is enough for calculating FO.
To first order in U, the modified Green function in real space is
\beqa
\label{eq:20}
G(\br,i\omega_n)&=&G^0(\br,i\omega)U G^0(-\br,i\omega_n)\nonumber \\
&=&\left(
                                                         \begin{array}{cc}
                                                           G_{cc}^0(\br,i\omega) & G_{cv}^0(\br,i\omega) \\
                                                           G_{vc}^0(\br,i\omega) & G_{vv}^0(\br,i\omega) \\
                                                         \end{array}
                                                       \right)\frac{u}{\sqrt{2}}\left(
                                                                \begin{array}{cc}
                                                                  1 & -i \\
                                                                  i & 1 \\
                                                                \end{array}
                                                              \right)\nonumber \\
                                                              &\times&\left(
                                                         \begin{array}{cc}
                                                           G_{cc}^0(-\br,i\omega) & G_{cv}^0(-\br,i\omega) \\
                                                           G_{vc}^0(-\br,i\omega) & G_{vv}^0(-\br,i\omega) \\
                                                         \end{array}
                                                       \right).
\eeqa
The Green function in $\bk$ space is
\beqa\label{eq:21}
    G^0(\bk,i\omega_n)&=&\left(
                                                              \begin{array}{cc}
                                                                G_{cc}^0(\bk,i\omega_n) & G_{cv}^0(\bk,i\omega_n) \\
                                                                G_{vc}^0(\bk,i\omega_n) & G_{vv}^0(\bk,i\omega_n) \\
                                                              \end{array}
                                                            \right)\nonumber\\&=&
    \frac{1}{i\omega_n-H_{kp}}.
\eeqa
For convenience, we denote
\beq
g\equiv\left[i\omega_n-\left(
                                     \begin{array}{cc}
                                       H_{cc} & 0 \\
                                       0 & H_{vv} \\
                                     \end{array}
                                   \right)\right]^{-1}
\eeq
 and take the non-diagonal part $\left(
                                   \begin{array}{cc}
                                     0 & \gamma k_x \\
                                     \gamma k_x & 0 \\
                                   \end{array}
                                 \right)
 $ as perturbation to get $G^0(\bk,i\omega_n)$. We will see the Hamiltonian can be reduced to a single band problem if we study FO with $E>E_c$.

Let the Fermi energy $E_F$ lie in the conduction band such that $k_F\ll\textrm{min}(1/a_x,1/a_y)$ and study the Friedel oscillation in BP. The $\bk$ space Green function $G_{cc}=(g_{cc}^{-1}+H_{cv}g_{vv}H_{vc})^{-1}$, $G_{cv}=g_{cc}H_{cv}G_{vv}$. Since ${\eta_v k_{x_F}^2+\nu_v k_{y_F}^2}\ll(E_c-E_v)$, we expand $H_{cv}g_{vv}H_{vc}$ in $1/(E_F-E_v)$, keeping terms up to quadratic in $\bk$, and obtain $G_{cc}^0({i\omega\rightarrow E+i\eta})=\frac{1}{E+i\eta-\eta_c k_x^2-\nu_c k_y^2-E_c -\frac{\gamma^2k_x^2}{E_c-E_v}}$. The non-diagonal part $G_{cv}$ is neglected in our calculation since $\gamma k_{x_F}/(E_F-E_v)\ll 1$. Written in real space,
\begin{eqnarray}\label{eq:22}
    G_{cc}^0(\br,E)&=&\frac{1}{N}\sum_\bk\frac{\exp{i\bk\cdot\br}}{E-\eta_c' k_x^2-\nu k_y^2-E_c}\nonumber \\
    &=&-i\frac{A_c}{2 \sqrt{\eta_c' \nu_c}}H_0^{(1)}\left(r'\sqrt{\frac{E-E_c}{\sqrt{\eta_c'\nu_c}}}\right)\nonumber\\
\end{eqnarray}
where ${\eta_c'\equiv \eta_c+\gamma^2/(E_c-E_v)}$, $r'^2\equiv\left(\sqrt{\frac{\nu_c}{\eta_c'}}x^2+\sqrt{\frac{\eta_c^\prime}{\nu_c}}y^2\right)$, and $H_0^{(1)}(z)$ is first-kind Hankel function. In this section, we denote $\textit{z}=r'\sqrt{\frac{E-E_c}{\sqrt{\eta_c'\nu_c}}}$ for convenience.

The impurity induced LDOS in real space is obtained from equation \eqref{eq:20}
\beqa
\delta\rho(E,\br)&=&-\frac{1}{\pi}\textrm{Im}[G(E,\br)U G(E,-\br)]_{cc}\nonumber\\
&=&G_{cc}(E,\br)\frac{u}{\sqrt{2}}G_{cc}(E,-\br)\nonumber\\
&=&\frac{\sqrt{2}u}{16\pi}A_c^2J_0(z)Y_0(z).
\eeqa

For large distance away from the impurity position, i.e. $z\gg 1$, keeping the leading terms in $J_0(z)Y_0(z)=-\cos(2z)/\pi z$ and integrate $\delta\rho(E,\bk)$ from the edge of conduction band to the Fermi energy, the leading part in FO induced by impurity in the large distance is
\begin{equation}\label{eq:4.3}
    \delta n(\br)=\frac{\sqrt{2}u}{16\pi}\frac{A_c^2}{\eta_c^\prime\nu_c}\frac{\sin\left(2\sqrt{\frac{E_F}{\eta_c^\prime}x^2+\frac{E_F}{\nu_c}y^2}\right)}{\frac{x^2}{\eta_c^\prime}+\frac{y^2}{\nu_c}}
\end{equation}
which is our main results in this sec..

From equation \eqref{eq:4.3}, it is seen that FO is anisotropic in BP. The oscillating part in the numerator is due to discontinuity at the Fermi surface. We conclude this section with comparing FO in BP to that in graphene. In graphene, the FO decay as $r^{-3}$ due to cancelation of modes on neighboring sites which decay as $r^{-2}$.~\cite{Virosztek2010} However, there is no such cancelation happening in BP because of the large gap. In BP, the FO shows similar behavior to ordinary 2-dimensional electron gas (2DEG) which shows $r^{-2}$ behavior. Due to different masses of carriers in $x$ and $y$ directions, FO in BP oscillates anisotropically in $x$ and $y$ direction.

\section{low density of impurities}
\label{sec:dos}

After the investigation of single-impurity problem,
we turn to many-impurity problem
in this section. For low density impurities which are short ranged and have finite amplitude, Born approximation is often sufficient. However, to study midgap states in band gap induced by impurities, Born approximation is not appropriate and we use $T$-matrix instead. Impurity is the main source of life time for carriers in the low temperature limit, therefore it is useful to calculate the impurities induced DOS. We first study the problem of impurities with $u\rightarrow\infty$, then the $t_4=0$ problem is considered. Next finite $v$ case is investigated. At last, we study impurities which have amplitude on all four sites for completeness.

Before detailed analysis of numerical results, it is worth understanding the origin of midgap states. For single impurity, it is seen from equation \eqref{eq:5} that impurity induced bound states comes from poles of $T_{\textrm{imp}}(E)$. For many impurities with small density, it will be seen midgap states are related to poles of $T(i\omega)$ in equation \eqref{eq:12}. Remember that for low density impurity, $T$-matrix approximation is justified and $T(E)=Un_i[1-U\overline{G}^0(E+i\eta)]^{-1}$. The DOS is
\beq
\label{eq:dos}
\rho_{X}(E)=-1/N\pi\rm{Im}[\sum_{\bk}G_{XX}(\bk,E+i\eta)]
\eeq
 with $X$ denotes A, B, C or D and $G(\bk,E+i\eta)$ is given in equation \eqref{eq:13}. If there is no impurity, BP has crystal symmetry $D_{2h}$ and DOS on A (B) is equal to C (D).~\cite{Ezawa2014}

The DOS is related to poles of Green function. Midgap states come from new poles associated with impurity. Expanding right hand side of equation \eqref{eq:13} to first order in $n_i$, i.e. $G=G^0+G^0T(i\omega)G^0$, it is seen that new poles come from $T(i\omega)$. The midgap states appear at $E$ satisfying
\begin{equation}\label{eq:24}
    \textrm{det}[I-V\sum_{\bk}G^0(\bk,E+i\eta)]=0
\end{equation}
just as bound states appear for a single impurity to fist order in $n_i$, we denote the solution of this equation as $E_{\textrm{imp}}$.
It is worth noting that our analysis is not suitable for long-range or high-density impurities where the interference effect is important.

\begin{figure}
  \includegraphics[width=9cm]{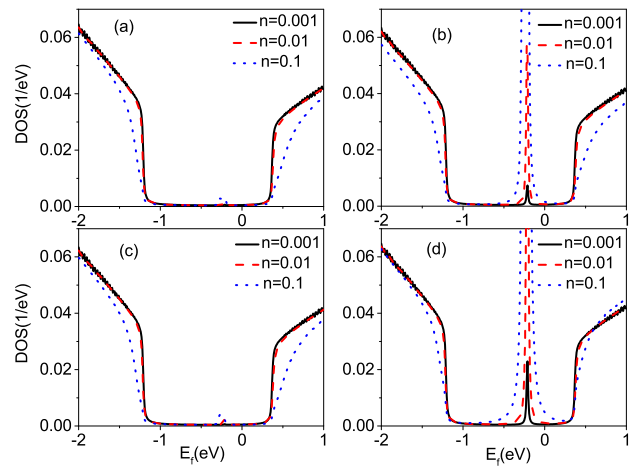}\\
  \caption{DOS on lattice A, B, C and D shown in figure (a), (b), (c) and (d) respectively in the case of $u\rightarrow\infty$. The DOS of midgap states on site A is very low as well as on C. DOS on B and D shown peaks in band gap which indicate midgap states. The impurity density $n$ is proportional to height of peaks.}\label{fig:dos1}
\end{figure}

We compute numerically the DOS in BP in the presence of impurities. The site dependent DOS is obtained from equation \eqref{eq:dos}. We first consider impurities of the form of equation \eqref{eq:4} and take the limit $u\rightarrow\infty$. As shown in figure \ref{fig:dos1}, we plot DOS on A, B, C and D site in BP separately. As we see in figure \ref{fig:dos1}, the DOS on A [figure \ref{fig:dos1}(a)] atom shows little evidence of bound states while DOS on B [figure \ref{fig:dos1}(b)] has midgap states. The limit $u\rightarrow\infty$ means decoupling the atom A from the system, the states belonging to the missing A atoms now have zero amplitude on site A and bound states may emerge. The DOS on C and D shows similar behavior to DOS on A and B respectively, the reason of which is that A (B) atom is connected to C (D) atom through small next nearest neighbor $t_4$ so the midgap state tend to stay on B and D atom.

To get more deep understanding of midgap states in BP, we note that the midgap states in BP are similar to midgap sates in graphene in the presence of vacancies. In graphene, that the midgap states in the presence of vacancies appear near $E=0$ eV is due to chiral symmetry of graphene Hamiltonian. Chiral symmetry in graphene is a symmetry is defined as $\sigma_3H\sigma_3=-H$ where $\sigma$ are Pauli matrices acting on pseudo-spin.~\cite{Wehling2009} Due to chiral symmetry, if there is a state at energy
$\varepsilon$, then there is also another state at energy $-\varepsilon$. For every $\bk$, there are two eigenstates for the graphene Hamiltonian $H(\bk)$ with $\varepsilon$ and $-\varepsilon$ respectively. If one atom is decoupled from this Hamiltonian, then only one state is allowed to exist and this state must appear at $\varepsilon=0$ eV.

\begin{figure}
  \includegraphics[width=9cm]{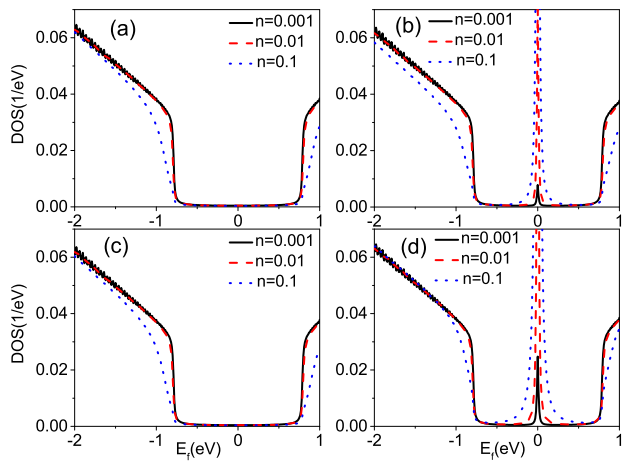}\\
  \caption{DOS on site A, B, C and D respectively with $t_4=0$ and $u\rightarrow\infty$. With $t_4=0$, BP preserves chiral symmetry due to which vacancies have bound states exactly at $E=0$ eV}\label{fig:dos2t4=0}
\end{figure}

Midgap states in BP can also be explained in this way because BP regains chiral symmetry if we set $t_4=0$ eV. To see it more clearly, we start from $2\times2$ Hamiltonian in reference [\onlinecite{Ezawa2014}] which reads
\begin{equation}\label{eq:25}
    H_{2\times2}=\left(
                   \begin{array}{cc}
                     f_{AC} & h \\
                     h^* & f_{AC} \\
                   \end{array}
                 \right),
\end{equation}
where $h$ is given in reference [~\onlinecite{Ezawa2014}] whose detailed expression is not relevant here.
The Hamiltonian \eqref{eq:25} has chiral symmetry if we set $t_4=0$ eV. So the midgap states should appear near $E=0$ eV if $t_4=0$ eV, we verify this by plotting DOS on A, B, C, and D sites respectively in BP where we have omitted $t_4$. The DOS is shown in figure \ref{fig:dos2t4=0}. As expected, midgap states appear near $E=0$ eV which is similar to graphene. So midgap states in BP given $t_4$ nonzero also should exist except they are shifted by an energy interval due to the nonzero $t_4$.

\begin{figure}
  \includegraphics[width=9cm]{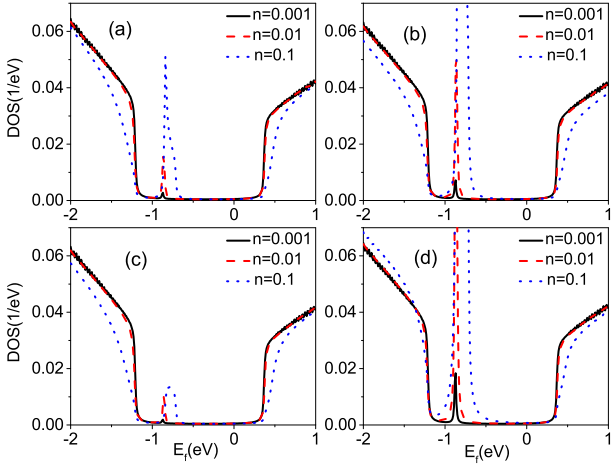}\\
  \caption{DOS on A, B, C and D respectively at $u=10$ eV. For finite $u$, DOS on A and C have finite amplitudes. The position of peaks move towards valence bands compared to $u\rightarrow\infty$.}\label{fig:dosv=10}
\end{figure}
\begin{figure}
  \includegraphics[width=9cm]{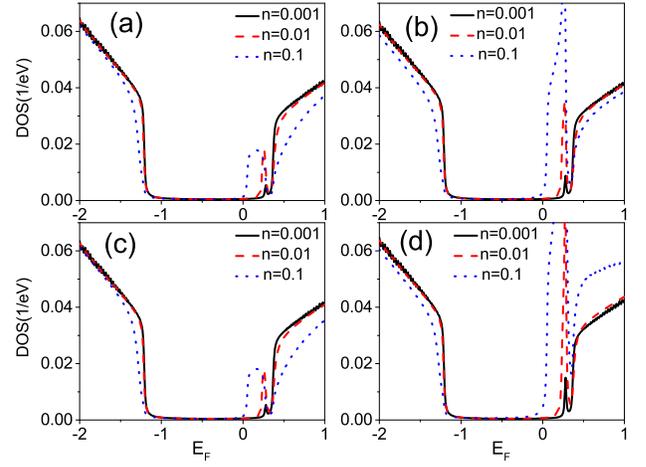}\\
 \caption{DOS on A, B, C and D respectively at $u=-10$ eV. For finite $u$, DOS on A and C have finite amplitudes. The position of peaks move towards conduction bands compared to $u\rightarrow\infty$.}\label{fig:dosv=-10}
\end{figure}
For finite $u$, the position $E_\textrm{imp}$ of midgap states shows different behaviors for positive $u$ and negative $u$. We plot $u=10$ eV and $-10$ eV in figure \ref{fig:dosv=10} and \ref{fig:dosv=-10} respectively. As already discussed before, impurity density does not affect position of midgap states though it may introduce new energy scales $nu$ and $nW$ with $W$ the width of band. Impurity density only affects heights of peaks. Position $E_{\rm{imp}}$ is determined by $u$. Positive $u$ intend to bound negative electrons or positive holes while negative $u$ attracts positive electrons. Note that there is a little difference between DOS on A and on C (also between B and D). The reason is that the presence of impurity at A break $D_{2h}$ symmetry and A is not equivalent to C in the presence of impurity which reside on site A.

\begin{figure}
  \includegraphics[width=6cm]{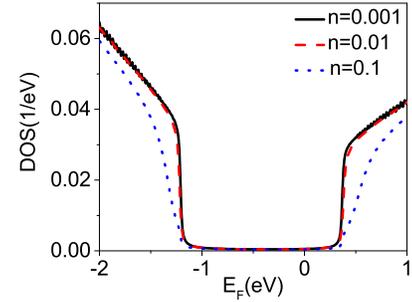}\\
  \caption{DOS on site A for impurity potential $u\rightarrow\infty$ with form of equation \eqref{eq:31} which has amplitudes on all four site in BP.The DOS on B, C and D are the same as A which we have not shown here. This kind impurity does not bound state in the band gap.}\label{figUI}
\end{figure}

For completeness we study the DOS in the presence of impurities which reside on all four sites, that is to say, impurity of this kind
\begin{equation}\label{eq:31}
    U=u\left(
         \begin{array}{cccc}
           1 & 0 & 0 & 0 \\
           0 & 1 & 0 & 0 \\
           0 & 0 & 1 & 0 \\
           0 & 0 & 0 & 1 \\
         \end{array}
       \right)
\end{equation}
 which has the same amplitude on all four sites. As shown in figure \ref{figUI} there is no midgap states in the gap if we take $u\rightarrow \infty$, the DOS on the four sites are the same so we present DOS on A only. The impurity reside on all four site in a unit cell so there is no site left to host a bound state with finite energy. For finite $u$, there is a solution $E_\textrm{imp}$ to equation \eqref{eq:24}. If $E_\textrm{imp}$ lies between the gap, midgap states will appear.~\cite{Wehling2007} On the other hand, if the solution $E_{\rm{imp}}$ lies in the bands, the impurity will modify the DOS in the bands and no midgap states is induced.~\cite{Katsnelson2015}

\begin{figure}
  \includegraphics[width=9cm]{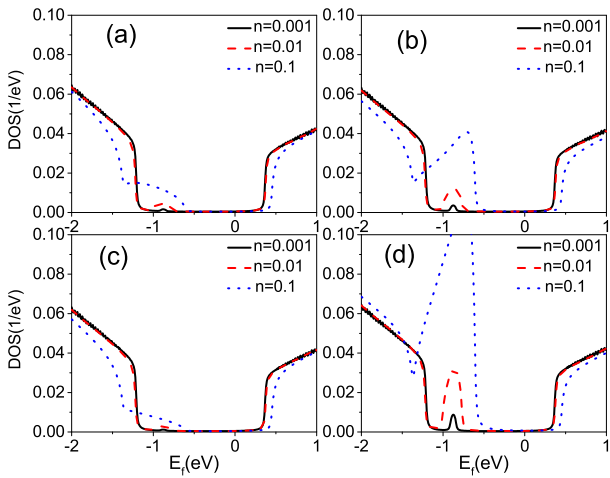}\\
  \caption{Self-consistent $T$-matrix calculation of impurity of the form in equation \eqref{eq:4}.  We set $u=10$ eV as in figure \ref{fig:dosv=10}.  Self-consistent calculation takes finite life time of carriers into account, so the width of peaks is larger than results using zeroth Green function.}\label{figsfct}
\end{figure}

At last self-consistent $T$-matrix is used to calculate DOS. Self-consistent $T$-matrix is also called full self-consistent Born approximation, it replaces the full Green function with zeroth order Green function in calculation of $T$-matrix. The same impurity problem as in figure \ref{fig:dosv=10} is calculated again using self-consistent $T$-matrix here. The numerical result is shown in figure \ref{figsfct}. There is no essential difference from $T$-matrix approximation which is shown in figure \ref{fig:dosv=10}, the position of $E_\textrm{imp}$ is not changed and amplitude is also proportional to height of peaks. But there are some differences indeed. The height and width of peaks have changed. It can be explained as follows, self-consistent calculation treats the imaginary parts of self energy more exactly, so the width of peaks in this approach is more reliable.

\section{Discussion and conclusions}
\label{sec:discuss}

We have discussed impurity problem in BP. For a single impurity we calculated FT-STS in momentum space and FO in real space. Numerical results of FT-STS is based on four band tight-binding Hamiltonian while FO is analytically calculated based on $2\times2\;\bk\cdot\bp$ Hamiltonian. The scattering is elastic so wave function will get phase shift after scattering. Due to phase shift, the interference between incoming and outgoing waves forms patterns in FT-STS. The largest part of the interference amplitude comes from backscattering in FT-STS. Because of anisotropy in BP, there are two features: one is that contours in FT-STS is elliptic, the other is that small contours appear in the inside of large contours. FO also shows anisotropy in oscillation and decaying directions. FO in BP is different from graphene in that it decays as $\frac{1}{x^2/\eta_c^\prime+y^2/\nu_c}$ while in graphene as $r^{-3}$.

For the many-impurity problem, we calculated DOS in BP and found that midgap states appear in the band gap. The position $E_{\rm{imp}}$ of midgap states is related to amplitude of impurity potential $u$. The density determines the height of peaks. The midgap states appear at $E=0$ for $u\rightarrow\infty$ when $t_4$ is set to zero, which is due to chiral symmetry of BP Hamiltonian in the absence of $t_4$. For finite $u$, the impurity site tends to bind negative (positive) charge carriers for positive (negative) $v$. We also calculated DOS using self-consistent $T$-matrix approximation and found the width of peaks is larger than DOS using $T$-matrix approximation.

In this work we have assumed the zero-temperature limit. We have omitted inelastic scattering due to phonons and electron-electron interactions. Indeed, at finite temperature, phonon mediated scattering is no longer energy conserved and it may have large effect on FT-STS. The states over a wide range of energy may take part into  scattering process and the phase space for scattering is enlarged compared to zero-temperature limit. So the phonon scattering may not be ignored in FT-STS at finite temperature. For electron-electron interaction, it has shown in graphene electron-electron interactions has strong influence on carriers near Dirac point in graphene.~\cite{Kotov2012} The Fermi velocity will be reshaped and even gap is opened at Dirac point due to exiton condensation.~\cite{Kotov2012} So it is expected that electron-electron may have influence on DOS in BP.

\begin{acknowledgments}
We thanks X. Y. Zhou and R. Zhang for many useful discussion, we also acknowledge for useful suggestions provided by H. Jiang, J. Liu and S. G. Cheng. This work was supported by Grant No. 2011CB922204 from the MOST of China and the National Natural Science Foundation (Grants No. 11474085, No. 11174252, No. 11304306 and No. 61290303).
\end{acknowledgments}

\end{document}